\documentclass{osa-article}

\usepackage{bm}
\usepackage{graphicx}
\usepackage{dcolumn}
\usepackage{textcomp}
\usepackage[utf8]{inputenc}
\usepackage[T1]{fontenc}
\usepackage{amsmath}
\usepackage{appendix}
\usepackage{color}
\usepackage{textcomp}
\usepackage{float}
\usepackage{lineno}
\allowdisplaybreaks[0]

\journal{osac}
\articletype{Research Article}

\begin{document}

\title{Enantio-detection via cavity-assisted three-photon processes}

\author{
Yu-Yuan Chen,\authormark{1} Chong Ye,\authormark{2,3} and Yong Li\authormark{1,4}
}

\address{
\authormark{1}Beijing Computational Science Research Center, Beijing 100193, China\\
\authormark{2}Beijing Key Laboratory of Nanophotonics and Ultrafine Optoelectronic Systems, School of Physics, Beijing Institute of Technology, 100081 Beijing, China\\
\authormark{3}yechong@bit.edu.cn\\
\authormark{4}liyong@csrc.ac.cn
}

\date{\today}

\begin{abstract}
We propose a method for enantio-detection of chiral molecules based on a cavity-molecules system, where the left- and right- handed molecules are coupled with a cavity and two classical light fields to form cyclic three-level models. Via the cavity-assisted three-photon processes based on the cyclic three-level model, photons are generated continuously in the cavity even in the absence of external driving to the cavity. However, the photonic fields generated from the three-photon processes of left- and right- handed molecules differ with the phase difference $\pi$ according to the inherent properties of electric-dipole transition moments of enantiomers. This provides a potential way to detect the enantiomeric excess of chiral mixture by monitoring the output field of the cavity.
\end{abstract}


\section{Introduction}
Molecular chirality has attracted much interest due to its fundamental role in the enantio-selective biological activity, chemical reactions, and pharmaceutical processes~\cite{EnantioEffect-Science1991,EnantioEffect-Nature1997,EnantioEffect-Drugs1996,EnantioEffect-Chiralitye2012}. Thus,  enantio-detection~\cite{Barron-MolecularScatter,Discrimination-CD,Discrimination-ChenYY,Discrimination-JSepSci2007,Discrimination-JiaWZ,Discrimination-Lehmann,Discrimination-Nature2000,Discrimination-OR,Discrimination-Science2002,Discrimination-VCD,Discrimination-XuXW,Discrimination-YeC,Discrimination-ZhangXD1,Discrimination-ZhangXD2}, spatial enantio-separation~\cite{Spatial-Separation-LiY-PRL,Spatial-Seperation-Andrews-NJP,Spatial-Seperation-Buhmann-PRA,Spatial-Seperation-Cameron-NJP,Spatial-Seperation-Hornberger-JCP,Spatial-Seperation-Shapiro-JCP,Spatial-Seperation-Shapiro-PRL,Spatial-Seperation-Cipparrone-LabChip,Spatial-Seperation-Tkachenko-NC1,Spatial-Seperation-Tkachenko-NC2,Spatial-Seperation-Zhao-NanoTechnol,Spatial-Seperation-Kravets-PRL,Spatial-Seperation-Cipparrone-LSAppl}, and inner-state enantio-purification (including the enantio-specific state transfer~\cite{Seperation-JiaWZ-JPB,Seperation-Koch-JCP,Seperation-LiY-PRA,Seperation-Schnell-ACIE,Seperation-SepDoyle-PRL,Seperation-Vitanov-PRA,Seperation-Vitanov-PRR,Seperation-YeC-PRA,Seperation-ZhangQS-JPB} and enantio-conversion~\cite{Conversion-Cohen-PRL,Conversion-Shapiro-JCP,Conversion-Shapiro-JPB,Conversion-Shapiro-PRL,Conversion-YeC-PRR,Conversion-YeC1,Conversion-YeC2,Conversion-Shapiro-PRA}) of chiral molecules are important and challenging tasks. Recently, there has been great success in the studies of enantio-detection of chiral molecules via the enantiomer-specific microwave spectroscopic methods~\cite{Microwave-Doyle-Nature,Microwave-Doyle-PCCP,Microwave-Doyle-PRL,Microwave-Lehmann-JPCL,Microwave-Schnell-ACIE,Microwave-Schnell-JPCL} based on the cyclic three-level models~\cite{ThreeLevel-Shapiro,ThreeLevel-LiuYX,ThreeLevel-Hirota,ThreeLevel-YeC,ThreeLevel-Vitanov,ThreeLevel-WuJL1,ThreeLevel-WuJL2}. In these methods, when the chiral molecules are driven by two classical light fields, a new classical light field (whose frequency is the sum of or difference between those of the two existed ones) is generated via the three-photon processes of three-wave mixing~\cite{Microwave-Doyle-Nature,Microwave-Doyle-PCCP,Microwave-Doyle-PRL,Microwave-Lehmann-JPCL,Microwave-Schnell-ACIE,Microwave-Schnell-JPCL}. Since the product of three electric-dipole transition moments for the cyclic three-level model changes sign with enantiomer~\cite{Microwave-Doyle-Nature,Microwave-Doyle-PRL}, there is a phase shift of $\pi$ between the new light fields generated from the left- and right- handed molecules. Thus, the difference between the numbers of left- and right- handed molecules is mapped on the total generated light field, and the enantiomeric excess of the chiral mixture can be detected by monitoring the total generated light field.

In the past few years, cavity-molecule(s) systems with a single molecule or many molecules confined in a cavity~\cite{CQEDMolecule1,CQEDMolecule2,CQEDMolecule3,CQEDMolecule4,CQEDMolecule5,CQEDMolecule6} have attained much attention. Recent investigations imply that there has been great progress in studying energy transfer~\cite{CQEDtransfer1,CQEDtransfer2}, control of chemical reactivity~\cite{CQEDReaction1,CQEDReaction2,CQEDReaction3}, and molecular spectra~\cite{CQEDMoleculeSpectra1,CQEDMoleculeSpectra2,CQEDMoleculeSpectra3,CQEDMoleculeSpectra4} in the cavity-molecule(s) systems. Such a system also provides a promising platform for the exploration of enantio-detection of chiral molecules~\cite{CQEDMolecule6}. Note that the realistic system for enantio-detection~\cite{Barron-MolecularScatter,Discrimination-CD,Discrimination-ChenYY, Discrimination-JSepSci2007,Discrimination-JiaWZ,Discrimination-Lehmann,Discrimination-Nature2000, Discrimination-OR,Discrimination-Science2002,Discrimination-VCD, Discrimination-XuXW,Discrimination-YeC,Discrimination-ZhangXD1,Discrimination-ZhangXD2} (or spatial enantio-separation~\cite{Spatial-Separation-LiY-PRL,Spatial-Seperation-Andrews-NJP,Spatial-Seperation-Buhmann-PRA,Spatial-Seperation-Cameron-NJP,Spatial-Seperation-Hornberger-JCP,Spatial-Seperation-Shapiro-JCP,Spatial-Seperation-Shapiro-PRL,Spatial-Seperation-Cipparrone-LabChip,Spatial-Seperation-Tkachenko-NC1,Spatial-Seperation-Tkachenko-NC2,Spatial-Seperation-Zhao-NanoTechnol,Spatial-Seperation-Kravets-PRL,Spatial-Seperation-Cipparrone-LSAppl} and inner-state enantio-purification~\cite{Seperation-JiaWZ-JPB,Seperation-Koch-JCP,Seperation-LiY-PRA,Seperation-Schnell-ACIE,Seperation-SepDoyle-PRL,Seperation-Vitanov-PRA,Seperation-Vitanov-PRR,Seperation-YeC-PRA,Seperation-ZhangQS-JPB,Conversion-Cohen-PRL,Conversion-Shapiro-JCP,Conversion-Shapiro-JPB,Conversion-Shapiro-PRL,Conversion-YeC-PRR,Conversion-YeC1,Conversion-YeC2,Conversion-Shapiro-PRA}) of chiral molecules usually involves a number of molecules. In the case of many molecules coupled with only the classical light fields~\cite{Barron-MolecularScatter,Discrimination-CD,Discrimination-ChenYY, Discrimination-JSepSci2007,Discrimination-JiaWZ,Discrimination-Lehmann, Discrimination-Nature2000,Discrimination-OR,Discrimination-Science2002, Discrimination-VCD,Discrimination-XuXW,Discrimination-YeC,Discrimination-ZhangXD1,Discrimination-ZhangXD2}, the single-molecule treatment describing the interaction between single molecules and light fields is appropriate. However, in the case that a quantized cavity field is introduced to couple with many molecules, the single-molecule treatment is usually no longer valid, instead the multi-molecule treatment is required to describe the collective interaction between the molecules and quantized cavity field~\cite{CQEDMolecule2,CQEDMolecule4,CQEDReaction1,CQEDtransfer2}.

In this paper, we propose a method for enantio-detection of chiral molecules via the cavity-assisted three-photon processes in the cavity-molecules system by resorting to the multi-molecule treatment, instead of taking the single-molecule one used in the previous methods for enantio-detection~\cite{Barron-MolecularScatter,Discrimination-CD,Discrimination-ChenYY,Discrimination-JSepSci2007,Discrimination-JiaWZ,Discrimination-Lehmann,Discrimination-Nature2000,Discrimination-OR,Discrimination-Science2002,Discrimination-VCD,Discrimination-XuXW,Discrimination-YeC,Discrimination-ZhangXD1,Discrimination-ZhangXD2,Microwave-Doyle-Nature,Microwave-Doyle-PCCP,Microwave-Doyle-PRL, Microwave-Lehmann-JPCL,Microwave-Schnell-ACIE,Microwave-Schnell-JPCL}. In our system, each molecule is modeled as a cyclic three-level model coupled by the quantized cavity field and two classical light fields. Initially, the chiral molecules stay in their ground states, and the quantized cavity field is in the vacuum state since there is no external driving to the cavity. When the two classical light fields are applied, the intracavity photons are generated continuously via the cavity-assisted three-photon processes, and will output from the cavity. Note that the photonic fields generated from the three-photon processes of left- and right- handed molecules differ with the phase difference $\pi$. Thus, the total generated photonic field is dependent on the difference between the numbers of left- and right- handed molecules in the chiral mixture. Based on this, we demonstrate the detection of enantiomeric excess via measuring the intensity of the cavity output field in the steady state.

Note that in the previous methods for enantio-detection, usually the single-molecule strong coupling (i.e., the coupling strength between single molecules and light field is much larger than the decay rates of molecules) is required to evade the influence of decoherence~\cite{Discrimination-JiaWZ,Discrimination-YeC,Discrimination-ChenYY}. In our method, however, such a condition of single-molecule strong coupling is not necessary since the involved collective interaction between many molecules and the quantized cavity field can be strong compared to the decay rates of molecules even the single-molecule coupling is weak~\cite{CQEDtransfer2,CQEDReaction1}. Moreover, in our cavity-molecules system, the intracavity photons are generated continuously via the cavity-assisted three-photon processes and output from the cavity along the direction of cavity axis. This implies that such a cavity-molecules system has potential advantages in the collection of the signal, compared with the previous enantiomer-specific microwave spectroscopic methods~\cite{Microwave-Doyle-Nature,Microwave-Doyle-PCCP,Microwave-Doyle-PRL,Microwave-Lehmann-JPCL,Microwave-Schnell-ACIE,Microwave-Schnell-JPCL} where the collection of the signal would be difficult. Therefore, our method offers an inspiring idea for the further exploration of enantio-detection.

This paper is organized as follows. In Sec.~\ref{ModEq}, we present the Hamiltonian and quantum Langevin equations for the cavity-molecules system under consideration. And the typical parameters used for discussions are also given. The results of enantio-detection in the case of low-excitation limit of molecules are given in Sec.~\ref{EmissionLow}. In Sec.~\ref{EmissionBeyondLow}, we further investigate the enantio-detection in the case beyond low-excitation limit of molecules. Finally, we summarize the conclusion in Sec.~\ref{summary}.

\section{Cavity-molecules system}\label{ModEq}

\begin{figure}[tbp]
	\centering
	\includegraphics[width=4.8cm]{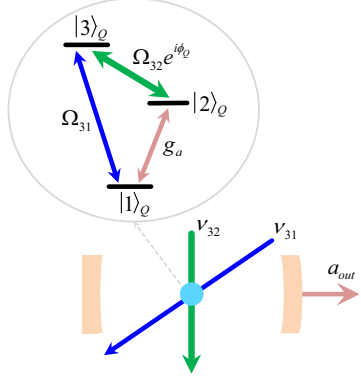}
	\caption{The schematics of the cavity-molecules system with the ensemble of $N_L$ left-handed molecules and $N_R$ right-handed ones confined in the (single-sided) cavity. Each molecule is coupled with the quantized cavity field (with angular frequency $\omega_{a}$) and two classical light fields (with angular frequencies $\nu_{31}$ and $\nu_{32}$, respectively) to form the cyclic three-level model, where the overall phase $\phi_Q$ is chirality dependent: $\phi_R=\phi_L+\pi$. }
	\label{Model}
\end{figure}

We now consider the cavity-molecules system consisting of a cavity (without external driving) and an ensemble of $N$ chiral molecules (confined in the cavity) as shown in Fig.~\ref{Model}. The ensemble contains $N_L$ left-handed and $N_R$ right-handed molecules. Each molecule is described by the cyclic three-level model, where the transition $\left|2\right\rangle_{Q} $$\leftrightarrow$$\left|1\right\rangle_{Q}$ with transition angular frequency $\omega_{21}$ is coupled to the quantized cavity field with angular frequency $\omega_{a}$. Meanwhile, the transition $\left|3\right\rangle_{Q}$$\leftrightarrow$$\left|1\right\rangle_{Q}$ ($\left|3\right\rangle_{Q}$$\leftrightarrow$$\left|2\right\rangle_{Q}$) with transition angular frequency $\omega_{31}$ ($\omega_{32}$) is coupled to a classical light field with angular frequency $\nu_{31}$ ($\nu_{32}$). The subscript $Q$ ($=L,R$) is introduced to denote the molecular chirality. Moreover, the intracavity photons are assumed to output only from the right side of the cavity for the sake of simplicity. By defining the collective operator for the chiral molecules $S_{jk}^{Q} = \sum_{m=1}^{N_{Q}} |j\rangle_{Q\, Q}^{m\, m} \hspace{-0.1em} \langle k|$ ($j,k=1,2,3$) and applying the dipole approximation and rotating-wave approximation, the Hamiltonian of the system reads ($\hbar=1$)
\begin{eqnarray}
H&=&\omega_{a}a^{\dagger}a + \omega_{21}(S_{22}^{L} + S_{22}^{R}) +\omega_{31}(S_{33}^{L} + S_{33}^{R}) +[g_a a (S_{21}^{L}+ S_{21}^{R})\nonumber\\
 & &+ \Omega_{31} e^{-i \nu_{31} t} (S_{31}^{L} + S_{31}^{R})
+\Omega_{32} e^{-i \nu_{32} t} (e^{i \phi_{L}} S_{32}^{L} + e^{i \phi_{R}} S_{32}^{R}) +\mathrm{H.c.}],
\label{HamiltonianO}
\end{eqnarray}
where $a$ ($a^\dag$) is the annihilation (creation) operator of the quantized cavity field, $g_{a}$ is the coupling strength between single molecules and the quantized cavity field, $\Omega_{31}$ and $\Omega_{32}$ are the coupling strengths between single molecules and the two classical light fields. These coupling strengths ($g_a$, $\Omega_{31}$, and $\Omega_{32}$) have been taken to be identical for all molecules by assuming the chiral mixture fixed in a volume whose size is much smaller than the typical wavelength. For simplicity but without loss of generality, they are taken as real. The chirality of the cyclic three-level model is specified by choosing the overall phases of left- and right- handed molecules as~\cite{Discrimination-JiaWZ,Discrimination-YeC,Spatial-Separation-LiY-PRL,Microwave-Doyle-Nature}
\begin{equation}
\phi_{L}=\phi,\ \ \phi_{R}=\phi+\pi.
\label{ChiralPhase}
\end{equation}

The Hamiltonian in the interaction picture with respect to ${H_0} = {(\nu_{31}-\nu_{32})}{a^\dag}a+(\nu_{31}-\nu_{32})(S_{22}^L + S_{22}^R) + {\nu_{31}}(S_{33}^L + S_{33}^R)$ can be written in the time-independent form as
\begin{eqnarray}
{H_I}&=&\,{\Delta_a}{a^\dag }a + (\Delta _{31} -\Delta _{32})(S_{22}^L +S_{22}^R) + {\Delta_{31}}(S_{33}^L + S_{33}^R)\nonumber\\
& &+[{g_a}a(S_{21}^L + S_{21}^R) + {\Omega_{31}}(S_{31}^L + S_{31}^R)
+ {\Omega_{32}}({e^{i{\phi_L}}}S_{32}^L + {e^{i{\phi_R}}}S_{32}^R) + {\rm{H.c.}}],
\label{HamiltonianA}
\end{eqnarray}
where $\Delta_a=\omega_{a}+\nu_{32}-\nu_{31}$, $\Delta_{31}=\omega_{31}-\nu_{31}$, and $\Delta_{32}=\omega_{32}-\nu_{32}$ are the detunings.

From Hamiltonian~(\ref{HamiltonianA}), we obtain the following quantum Langevin equations~\cite{QuantumOpticsBook1,QuantumNoise}
\begin{eqnarray}
\dot{a}&=&-i\Delta_a a -i g_{a}(S_{12}^{L} + S_{12}^{R}) -{\kappa_a}a +\sqrt{2\kappa_{a}}a_{\rm{in}}, \nonumber\\
\dot{S}_{12}^{Q}&=& i[(\Delta_{32}-\Delta_{31})S_{12}^{Q} + g_a a (S_{22}^{Q}-S_{11}^{Q}) + \Omega_{31}S_{32}^{Q} - \Omega_{32}e^{-i\phi_{Q}}S_{13}^{Q}] - \Gamma_{21}S_{12}^{Q}+F_{12}^{Q}, \nonumber\\
\dot{S}_{13}^{Q}&=& i[-\Delta_{31}S_{13}^{Q} + \Omega_{31}(S_{33}^{Q}-S_{11}^{Q}) + g_a a S_{23}^{Q} - \Omega_{32}e^{i\phi_{Q}}S_{12}^{Q}] - (\Gamma_{31}+\Gamma_{32})S_{13}^{Q}+F_{13}^{Q}, \nonumber\\
\dot{S}_{23}^{Q}&=& i[-\Delta_{32}S_{23}^{Q} + \Omega_{32}e^{i\phi_{Q}}(S_{33}^{Q}-S_{22}^{Q}) - \Omega_{31}S_{21}^{Q} +g_a a^{\dagger}S_{13}^{Q}] - (\Gamma_{21}+\Gamma_{31}+\Gamma_{32})S_{23}^{Q}+F_{23}^{Q}, \nonumber\\
\dot{S}_{11}^{Q}&=& i [g_a(aS_{21}^{Q}-a^{\dagger}S_{12}^{Q})+\Omega_{31}(S_{31}^{Q}-S_{13}^{Q})]+2\Gamma_{21}S_{22}^{Q}+2\Gamma_{31}S_{33}^{Q}+F_{11}^{Q}, \nonumber\\
\dot{S}_{22}^{Q}&=& i [g_a(a^{\dagger}S_{12}^{Q}-aS_{21}^{Q})
+\Omega_{32}(e^{i\phi_{Q}}S_{32}^{Q}-e^{-i\phi_{Q}}S_{23}^{Q})]-2\Gamma_{21}S_{22}^{Q}+2\Gamma_{32}S_{33}^{Q}+F_{22}^{Q}.
\end{eqnarray}
Here, $\kappa_{a}$ represents the photonic decay rate from the right side of the (single-sided) cavity and we have neglected the other photonic decay rates. $a_{\rm{in}}$ denotes the quantum input noise operator of the cavity with zero-mean value (i.e., $\langle a_{\rm{in}} \rangle=0$). $\Gamma_{21}$, $\Gamma_{31}$, and $\Gamma_{32}$ represent the decay rates of the molecules associated with the three transitions, respectively. $F_{jk}^{Q}$ (with $j,k=1,2,3$) denotes the quantum noise term for the collective operator $S_{jk}^{Q}$, and has zero-mean value (i.e., $\langle F_{jk}^{Q} \rangle=0$). We here consider only the decay rates of molecules resulting from pure population relaxation of collisional processes and have assumed the pure dephasing collisional effects are negligible. This could be justified for the case of molecules in a buffer gas cooling cell~\cite{decoherence-Doyle-2012,Microwave-Doyle-Nature,Microwave-Doyle-PRL,Microwave-Doyle-PCCP}. Taking the mean-field approximation~\cite{QuantumOpticsBook2}, we obtain the corresponding steady-state equations as
\begin{eqnarray}
0&=&-(i\Delta_a+{\kappa_a}) \langle a \rangle -i g_{a}(\langle S_{12}^{L} \rangle +\langle S_{12}^{R} \rangle), \nonumber\\
0&=&-[i(\Delta_{31}-\Delta_{32})+\Gamma_{21}]\langle S_{12}^{Q}\rangle + ig_a \langle a\rangle(\langle S_{22}^{Q}\rangle - \langle S_{11}^{Q}\rangle) + i\Omega_{31}\langle S_{32}^{Q}\rangle - i\Omega_{32}e^{-i\phi_{Q}}\langle S_{13}^{Q}\rangle, \nonumber\\
0&=&-[i\Delta_{31}+(\Gamma_{31}+\Gamma_{32})]\langle S_{13}^{Q}\rangle + i\Omega_{31}(\langle S_{33}^{Q}\rangle - \langle S_{11}^{Q}\rangle) + ig_a \langle a\rangle\langle S_{23}^{Q}\rangle - i\Omega_{32}e^{i\phi_{Q}}\langle S_{12}^{Q}\rangle, \nonumber\\
0&=&-[i\Delta_{32}+(\Gamma_{21}+\Gamma_{31}+\Gamma_{32})]\langle S_{23}^{Q}\rangle + i\Omega_{32}e^{i\phi_{Q}}(\langle S_{33}^{Q}\rangle - \langle S_{22}^{Q}\rangle) - i\Omega_{31} \langle S_{21}^{Q}\rangle + ig_a \langle a^{\dagger}\rangle\langle S_{13}^{Q}\rangle, \nonumber\\
0&=& ig_a(\langle a\rangle\langle S_{21}^{Q}\rangle - \langle a^{\dagger}\rangle\langle S_{12}^{Q}\rangle) + i\Omega_{31}(\langle S_{31}^{Q}\rangle - \langle S_{13}^{Q}\rangle) + 2\Gamma_{21}\langle S_{22}^{Q}\rangle + 2\Gamma_{31}\langle S_{33}^{Q}\rangle, \nonumber\\
0&=& ig_a(\langle a^{\dagger}\rangle\langle S_{12}^{Q}\rangle - \langle a\rangle\langle S_{21}^{Q}\rangle) + i\Omega_{32}(e^{i\phi_{Q}}\langle S_{32}^{Q}\rangle - e^{-i\phi_{Q}}\langle S_{23}^{Q}\rangle) - 2\Gamma_{21}\langle S_{22}^{Q}\rangle + 2\Gamma_{32}\langle S_{33}^{Q}\rangle,
\label{SteadyLangevinNLow}
\end{eqnarray}
where $\langle O \rangle$ (with $O=a,\,a^{\dagger},\,S_{jk}^{Q}$) denotes the mean value of the operator $O$. Note that the mean values $\langle S_{jj}^{Q}\rangle$ (with $j=1,2,3$) satisfy the relation $\sum_{j=1}^3 \langle S_{jj}^{Q} \rangle = N_Q$.

In the following discussions, we take 1,2-propanediol~\cite{PropanediolParameter1994,PropanediolParameter2017} as an example to demonstrate our method. The working states of the cyclic three-level model for the chiral molecules are selected as $|1\rangle=|g\rangle\left|0_{000}\right\rangle$, $|2\rangle=|e\rangle\left|1_{110}\right\rangle$, and $|3\rangle=|e\rangle\left|1_{101}\right\rangle$, where $|g\rangle$ and $|e\rangle$ denote, respectively, the vibrational ground and first-excited states for the motion of OH-stretch with the transition angular frequency $\omega_{\text{vib}}=2\pi\times100.95\,\text{THz}$~\cite{PropanediolParameter1994}. The rotational states are marked in the $\left|J_{K_{a}K_{c}M}\right\rangle$ notation~\cite{Seperation-Koch-JCP,anglemomentum}. $J$ is the angular momentum quantum number, $K_a$ ($K_c$) runs from $J$ ($0$) to $0$ ($J$) in unit step with the decrease of energy, and $M$ denotes the magnetic quantum number. Based on the rotational constants for 1,2-propanediol $A=2\pi\times8524.405\,\text{MHz}$, $B=2\pi\times3635.492\,\text{MHz}$, and $C=2\pi\times2788.699\,\text{MHz}$~\cite{PropanediolParameter2017}, the transition angular frequencies are calculated as $\omega_{21}=2\pi\times100.961\,\text{THz}$, $\omega_{31}=2\pi\times100.962\,\text{THz}$, and $\omega_{32}=2\pi\times0.8468\,\text{GHz}$~\cite{anglemomentum}.

\section{Enantio-detection in the low-excitation limit of molecules}\label{EmissionLow}

In this section, we investigate the steady-state output field of the cavity in the case of low-excitation limit of molecules and explore its potential applications in the detection of enantiomeric excess. Here, we consider the case of weak coupling strength $\Omega_{31} \ll \Gamma_{31}$ since such a weak coupling strength can usually ensure the low-excitation limit of molecules, that is, most molecules stay in the ground state $\left|1\right\rangle$ and only a few of them will be excited. At the end of this section, we will verify the validity of such a limit for the parameters used.

\subsection{Steady-state intensity of the output field}

For the case of weak coupling strength $\Omega_{31} \ll \Gamma_{31}$, one can treat $\Omega_{31}$ as a perturbation parameter. Hence, we apply the perturbation approach to the element $\langle O\rangle$, which is given in terms of perturbation expansion~\cite{Perturbation} with respect to $\Omega_{31}$
\begin{equation}
\langle O\rangle=\langle O\rangle^{(0)}+\langle O\rangle^{(1)}+\langle O\rangle^{(2)}+\cdots, \label{perturbation}
\end{equation}
where $\langle O\rangle^{(n)}~(n=0,1,...)$ is the $n$-th-order solution of $\langle O\rangle$. The zeroth-order solution $\langle O\rangle^{(0)}$ corresponds to the case in the absence of the classical light field $\Omega_{31}$. In such a case, no photons are generated, and thus the molecules in the initial ground state $\left|1\right\rangle$ cannot be excited to the excited states $\left|2\right\rangle$ and $\left|3\right\rangle$. Therefore, the zeroth-order solutions are given by
\begin{eqnarray}
&&\langle a\rangle^{(0)}=0,~\langle S_{11}^{Q}\rangle^{(0)} = N_{Q}, \nonumber\\
&&\langle S_{22}^{Q}\rangle^{(0)}=\langle S_{33}^{Q}\rangle^{(0)} = 0,~\langle S_{jk}^{Q}\rangle^{(0)}=0~(k\neq j).
\label{Zero-orderSolutions}
\end{eqnarray}
In Eq.~(\ref{Zero-orderSolutions}), we have assumed that the system is in the ideal zero-temperature environment, where there are no photons in the cavity (i.e., $\langle a\rangle^{(0)}=0$) and all the molecules stay in the ground state $\left|1\right\rangle$ (i.e., $\langle S_{11}^{Q}\rangle^{(0)} = N_{Q}$). This is reasonable since for the cavity with infrared angular frequency $\omega_{a}\sim 2\pi\times100.961\,\text{THz}$, one has $\hbar\omega_{a} \gg k_B T$ even in the room-temperature environment ($T \sim 300\,\text{K}$) with $k_B$ the Boltzmann constant. Meanwhile, for the molecules with infrared transition angular frequency (e.g., $\omega_{21}=2\pi\times100.961\,\text{THz}$ and $\omega_{31}=2\pi\times100.962\,\text{THz}$ for 1,2-propanediol molecules) and in typical cooled vibrational temperature $T_{\text{vib}}\simeq100\,\text{K}$~\cite{decoherence-Doyle-2012}, we find $\{\hbar\omega_{21},\,\hbar\omega_{31}\}\gg k_B T_{\text{vib}}$. Therefore, we can neglect the effect of noises on the steady-state solution $\langle O\rangle^{(0)}$ safely~\cite{QuantumNoise}.

Substituting Eq.~(\ref{perturbation}) into Eq.~(\ref{SteadyLangevinNLow}), one can find that the first-order solutions satisfy the following equations
\begin{eqnarray}
0&=&-(i\Delta_a+{\kappa_a}) \langle a \rangle^{(1)} -i g_{a}[\langle S_{12}^{L} \rangle^{(1)} +\langle S_{12}^{R} \rangle^{(1)}], \nonumber\\
0&=&-[i(\Delta_{31}-\Delta_{32})+\Gamma_{21}]\langle S_{12}^{Q}\rangle^{(1)} + ig_a \langle a\rangle^{(1)}[\langle S_{22}^{Q}\rangle^{(0)} - \langle S_{11}^{Q}\rangle^{(0)}]\nonumber\\
&&+ ig_a \langle a\rangle^{(0)}[\langle S_{22}^{Q}\rangle^{(1)} - \langle S_{11}^{Q}\rangle^{(1)}]+i\Omega_{31}\langle S_{32}^{Q}\rangle^{(0)}- i\Omega_{32}e^{-i\phi_{Q}}\langle S_{13}^{Q}\rangle^{(1)}, \nonumber\\
0&=&-[i\Delta_{31}+(\Gamma_{31}+\Gamma_{32})]\langle S_{13}^{Q}\rangle^{(1)} + i\Omega_{31}[\langle S_{33}^{Q}\rangle^{(0)} - \langle S_{11}^{Q}\rangle^{(0)}] + ig_a \langle a\rangle^{(1)} \langle S_{23}^{Q}\rangle^{(0)} \nonumber\\
&&+ ig_a \langle a\rangle^{(0)} \langle S_{23}^{Q}\rangle^{(1)} - i\Omega_{32}e^{i\phi_{Q}}\langle S_{12}^{Q}\rangle^{(1)}, \nonumber\\
0&=&-[i\Delta_{32}+(\Gamma_{21}+\Gamma_{31}+\Gamma_{32})]\langle S_{23}^{Q}\rangle^{(1)} + i\Omega_{32}e^{i\phi_{Q}}[\langle S_{33}^{Q}\rangle^{(1)} - \langle S_{22}^{Q}\rangle^{(1)}] - i\Omega_{31} \langle S_{21}^{Q}\rangle^{(0)} \nonumber\\
&&+ig_a \langle a^{\dagger}\rangle^{(1)} \langle S_{13}^{Q}\rangle^{(0)} + ig_a \langle a^{\dagger}\rangle^{(0)} \langle S_{13}^{Q}\rangle^{(1)}, \nonumber\\
0&=& ig_a[\langle a\rangle^{(1)} \langle S_{21}^{Q}\rangle^{(0)} +\langle a\rangle^{(0)} \langle S_{21}^{Q}\rangle^{(1)} - \langle a^{\dagger}\rangle^{(1)} \langle S_{12}^{Q}\rangle^{(0)} - \langle a^{\dagger}\rangle^{(0)} \langle S_{12}^{Q}\rangle^{(1)}] \nonumber\\
&&+ i\Omega_{31}[\langle S_{31}^{Q}\rangle^{(0)} - \langle S_{13}^{Q}\rangle^{(0)}] + 2\Gamma_{21}\langle S_{22}^{Q}\rangle^{(1)} + 2\Gamma_{31}\langle S_{33}^{Q}\rangle^{(1)}, \nonumber\\
0&=& ig_a[\langle a^{\dagger}\rangle^{(1)} \langle S_{12}^{Q}\rangle^{(0)} + \langle a^{\dagger}\rangle^{(0)} \langle S_{12}^{Q}\rangle^{(1)} - \langle a\rangle^{(1)} \langle S_{21}^{Q}\rangle^{(0)} -\langle a\rangle^{(0)} \langle S_{21}^{Q}\rangle^{(1)}] \nonumber\\
&&+i\Omega_{32}[e^{i\phi_{Q}}\langle S_{32}^{Q}\rangle^{(1)} - e^{-i\phi_{Q}}\langle S_{23}^{Q}\rangle^{(1)}]- 2\Gamma_{21}\langle S_{22}^{Q}\rangle^{(1)}+ 2\Gamma_{32}\langle S_{33}^{Q}\rangle^{(1)}, \nonumber\\
0&=&\langle S_{11}^{Q} \rangle^{(1)} + \langle S_{22}^{Q} \rangle^{(1)} +\langle S_{33}^{Q} \rangle^{(1)}.
\label{First-orderEquations}
\end{eqnarray}
From Eqs.~(\ref{ChiralPhase}),~(\ref{Zero-orderSolutions}),~and~(\ref{First-orderEquations}), we get the first-order solution of $\langle a\rangle$ as
\begin{equation}
\langle a\rangle^{(1)}=\frac{i (N_{L}-N_{R}) g_a \Omega_{31} \Omega_{32} e^{-i \phi}} {K_a(K_{21} K_{31}+\Omega_{32}^{2})+g_a^{2}N K_{31}},
\label{OutputField}
\end{equation}
where $N=N_L+N_R$, $K_a=i\Delta_a+{\kappa_a}$, $K_{21}=i(\Delta_{31}-\Delta_{32}) + \Gamma_{21}$, and $K_{31}=i\Delta_{31} + \Gamma_{31} + \Gamma_{32}$. For the case of weak coupling strength $\Omega_{31}$, the first-order solution $\langle a\rangle^{(1)}$ sufficiently reflects the main physical properties of $\langle a\rangle$. Therefore, here we take the approximation $\langle a\rangle \simeq \langle a\rangle^{(1)}$.

In our system, the intracavity photons can only output from the right side of the cavity. Using the input-output relation~\cite{QuantumNoise,QuantumOpticsBook3}
\begin{equation}
\sqrt{2\kappa_{a}} a  = a_{\rm{in}}+a_{\rm{out}},
\label{InOutput}
\end{equation}
we obtain the mean output field from the right side of the cavity $\left\langle a_{\rm{out}}\right\rangle = \sqrt{2\kappa_{a}} \left\langle a \right\rangle \simeq \sqrt{2\kappa_{a}}  \langle a\rangle^{(1)} $. It is obvious that such a mean output field depends on the difference between the numbers of left- and right- handed molecules (i.e., $N_{L}-N_{R}$) according to Eq.~(\ref{OutputField}). The underlying physics is as follows. When the chiral mixture confined in the cavity is coupled to the two classical light fields, the cavity-assisted three-photon processes [seen from the term $g_a \Omega_{31} \Omega_{32}$ in the numerator of Eq.~(\ref{OutputField})] lead to the continuous generation of intracavity photons. Since $g_a \Omega_{31} \Omega_{32} e^{-i \phi_{Q}}$ changes sign with enantiomer, the total contributions of the cavity-assisted three-photon processes of the chiral mixture to the output field of the cavity is dependent on $N_{L}-N_{R}$.

According to the definition $I_{\rm{out}}=\left| \left\langle a_{\rm{out}}\right\rangle \right|^2$, one further obtains the intensity of the output field as
\begin{equation}
I_{\rm{out}} =\bigg| \frac{N \sqrt{2\kappa_a} g_a \Omega_{31} \Omega_{32} \eta } {K_a(K_{21} K_{31}+\Omega_{32}^{2})+g_a^{2}N K_{31}} \bigg|^{2},
\label{Output}
\end{equation}
where $\eta=(N_{L}-N_{R})/N$ denotes the enantiomeric excess. From Eq.~(\ref{Output}), one can find that the intensity of the output field $I_{\rm{out}}$ does not depend on the overall phase $\phi$.

In the previous enantio-detection methods involving only the interaction between many molecules and classical light fields~\cite{Discrimination-JiaWZ,Discrimination-YeC,Discrimination-ChenYY}, the single-molecule strong coupling is usually required to evade the influence of decoherence. In our method involving the collective interaction between many molecules and the quantized cavity field, the presence of the quantized cavity field enables the collectively enhanced coupling~\cite{CQEDtransfer2,CQEDReaction1,CollectiveCoupling1,CollectiveCoupling2,CollectiveCoupling3,CollectiveCoupling4} $g_{a}\sqrt{N} \gg \{\Gamma_{jk},\,\kappa_{a}\}$ even for the case of weak single-molecule coupling strength $g_{a} \lesssim \{ \Gamma_{jk},\,\kappa_{a} \}$. 
In the following discussions, we focus on such a region where the single-molecule coupling strength is weak but the collective coupling strength is strong compared with the decay rates of the cavity and molecules.

\subsection{Detection of the enantiomeric excess}

So far, we have obtained the intensity of the output field resulting from the cavity-assisted three-photon processes for the chiral mixture. In the following analysis and simulations, the transition $\left|2\right\rangle_{Q} $$\leftrightarrow$$\left|1\right\rangle_{Q}$ ($\left|3\right\rangle_{Q} $$\leftrightarrow$$\left|1\right\rangle_{Q}$) is assumed to be resonantly coupled with the quantized cavity field (classical light field), which means $\Delta_{a}=-\Delta_{32}$ ($\Delta_{31}=0$).

\begin{figure}[ht]
\centering
\includegraphics[width=8.5cm]{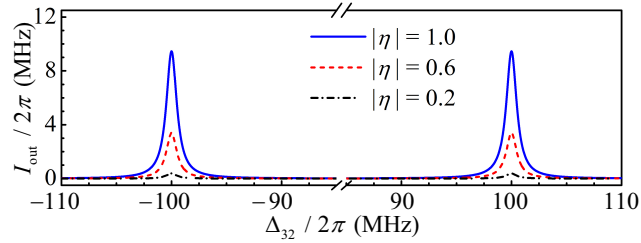}
\caption{The intensity of the output field as a function of the detuning $\Delta_{32}$ for different enantiomeric excess $\eta$. The parameters are $\Gamma_{21}/2\pi=\Gamma_{32}/2\pi=\Gamma_{31}/2\pi=0.1\,\text{MHz}$, $\kappa_a/2\pi=1\,\text{MHz}$, $\Omega_{31}/2\pi=5\,\text{kHz}$, $\Omega_{32}/2\pi=0.1\,\text{MHz}$, $g_{a}/2\pi=0.1\,\text{MHz}$, $\Delta_{31}=0$, $\Delta_{a}=-\Delta_{32}$, and $N=10^6$.}
\label{eeDepenLOW}
\end{figure}

In Fig.~\ref{eeDepenLOW}, we give the intensity of the output field $I_{\rm{out}}$ versus the detuning $\Delta_{32}$ for different enantiomeric excess $\eta$. Here, we take the weak coupling strength $\Omega_{31}/2\pi=5\,\text{kHz}$. We choose the typical available parameters: the total number of chiral molecules $N=10^6$~\cite{CQEDMolecule4,CQEDReaction1}, the coupling strength $g_a/2\pi=0.1\,\text{MHz}$~\cite{CavityCoupling-Long-2015,CavityCoupling-Shalabney-2015,CavityCoupling-Vergauwe-2016,CavityCouplingDecay-Kampschulte-2018}, and the photonic decay rate from the cavity $\kappa_a/2\pi=1\,\text{MHz}$~\cite{CavityCouplingDecay-Kampschulte-2018,CavityDecay-Hua-2018,CavityDecay-Hoghooghi-2019}. For simplicity, we take the decay rates of molecules $\Gamma_{21}/2\pi=\Gamma_{32}/2\pi=\Gamma_{31}/2\pi=0.1\,\text{MHz}$ according to typical experimental parameters~\cite{Microwave-Doyle-Nature,Microwave-Doyle-PRL}. It is shown that there exist two characteristic peaks at the detuning $\Delta_{32} \simeq \pm g_{a}\sqrt{N}$. This can be understood as the result of the vacuum Rabi splitting induced by the quantized cavity field in the strong collective coupling condition~\cite{QuantumOpticsBook2,RabiSplit1,RabiSplit2,RabiSplit3,RabiSplit4,CQEDtransfer2,CQEDReaction1}. Around the two characteristic peaks at $\Delta_{32} \simeq \pm g_{a}\sqrt{N}$, $I_{\rm{out}}$ is relatively sensitive to $\eta$ compared with those at other detunings. 

\begin{figure}[ht]
	\centering
	\includegraphics[width=8.5cm]{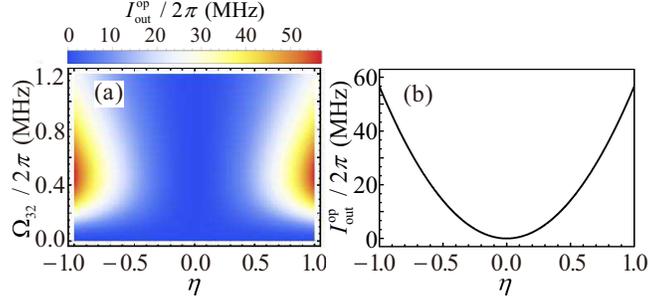}
	\caption{(a) The optimal cavity output intensity $I_{\rm{out}}^{\rm{op}}$ versus the enantiomeric excess $\eta$ and the coupling strength $\Omega_{32}$. (b) The optimal cavity output intensity $I_{\rm{out}}^{\rm{op}}$ versus the enantiomeric excess $\eta$ for $\Omega_{32}/2\pi=0.5\,\text{MHz}$. The other parameters are the same as those in Fig.~\ref{eeDepenLOW} except $\Delta_{32} = g_a \sqrt{N}$.}
	\label{eeDetectLOW}
\end{figure}

Here we are interested in the optimal cavity output intensity at $\Delta_{32} =g_{a}\sqrt{N}$:
\begin{equation}
I_{\rm{out}}^{\rm{op}} \simeq  \frac{2 N \kappa_a \Omega_{31}^2\Omega_{32}^2{\eta}^2}{\left[(\Gamma_{31}+\Gamma_{32})(\Gamma_{21}+\kappa_{a})+\Omega_{32}^2\right]^2},
\label{Peak}
\end{equation}
which is obtained by substituting $\Delta_{32} =g_{a}\sqrt{N}$ into Eq.~(\ref{Output}) and considering the strong collective coupling condition $g_{a}\sqrt{N} \gg \{ \Gamma_{jk},\,\kappa_{a} \}$. The optimal cavity output intensity $I_{\rm{out}}^{\rm{op}}$ versus the enantiomeric excess $\eta$ and the coupling strength $\Omega_{32}$ is displayed in Fig.~\ref{eeDetectLOW}(a). It is shown that $I_{\rm{out}}^{\rm{op}}$ strongly depends on the coupling strength $\Omega_{32}$. When $\Omega_{32}/2\pi \simeq 0.5\,\text{MHz}$, the optimal cavity output intensity for $|\eta|=1$ reaches the maximal $I_{\rm{out}}^{\rm{op}}/2\pi \simeq 57\,\text{MHz}$ [see Fig.~\ref{eeDetectLOW}(b)]. Seen from Fig.~\ref{eeDetectLOW}(b) or Eq.~(\ref{Output}), it is clear that one can detect the enantiomeric excess of chiral mixture by measuring the intensity of the output field.

We would like to remark that the above discussions are based on the low-excitation limit of molecules in the case of weak coupling strength $\Omega_{31}$. Thus, we should verify whether they meet the requirement for such a limit. Here, the factor
\begin{equation}
P_e^{Q}=\frac{\langle S_{22}^{Q}\rangle+\langle S_{33}^{Q}\rangle}{N_{Q}}
\end{equation}
is introduced to describe the proportion of the left- (right-) handed molecules occupying the excited states $\left|2\right\rangle_L$ and $\left|3\right\rangle_L$ ($\left|2\right\rangle_R$ and $\left|3\right\rangle_R$) to the total left- (right-) handed molecules. The steady-state values $\langle S_{22}^{Q} \rangle$ and $\langle S_{33}^{Q} \rangle$ are obtained by numerically solving Eq.~(\ref{SteadyLangevinNLow}). For the parameters used in Figs.~\ref{eeDepenLOW} and~\ref{eeDetectLOW}, we find $P_e^{Q} < 0.3\%$. This implies that most molecules occupy the ground state $\left|1\right\rangle$, meeting the requirement of the low-excitation limit of molecules in the above discussions.

\section{Enantio-detection beyond low-excitation limit of molecules}\label{EmissionBeyondLow}

In Sec.~\ref{EmissionLow}, we have shown the achievement of enantio-detection by investigating the steady-state output field of the cavity in the case of low-excitation limit of molecules for weak coupling strength $\Omega_{31}$. In this section, we further study the steady-state output field of the cavity in the case beyond low-excitation limit of molecules by increasing the coupling strength $\Omega_{31}$, and show the related enantio-detection is also possible.

\subsection{Steady-state intensity of the output field}

In the case beyond low-excitation limit of molecules, one can obtain the steady-state values ($\left\langle a \right\rangle$, $\langle a^{\dagger} \rangle$, and $\langle S_{jk}^Q \rangle$) by numerically solving the steady-state equations~(\ref{SteadyLangevinNLow}). Using the input-output relation (\ref{InOutput}), we can also obtain the mean output field $\left\langle a_{\rm{out}}\right\rangle = \sqrt{2\kappa_{a}} \left\langle a \right\rangle$ and its intensity $I_{\rm{out}}=\left| \left\langle a_{\rm{out}}\right\rangle \right|^2$.

In the following analysis and simulations, we still focus on the region where the single-molecule coupling strength is weak $g_{a} \lesssim \{\Gamma_{jk},\,\kappa_{a}\}$ but the strong collective coupling condition~\cite{CollectiveCoupling1,CollectiveCoupling2,CollectiveCoupling4} $g_{a}\sqrt{N} \gg \{\Gamma_{jk},\,\kappa_{a}\}$ is satisfied. The detunings ($\Delta_a$ and $\Delta_{31}$), the total number of chiral molecules $N$, the coupling strength $g_a$, and the decay rates ($\kappa_a$, $\Gamma_{21}$, $\Gamma_{31}$, and $\Gamma_{32}$) are taken as the same values as those in the case of low-excitation limit of molecules in Sec.~\ref{EmissionLow}. Moreover, we choose the coupling strength $\Omega_{32}/2\pi=0.5\,\text{MHz}$.

\subsection{Detection of the enantiomeric excess}

\begin{figure}[ht]
	\centering
	\includegraphics[width=8.6cm]{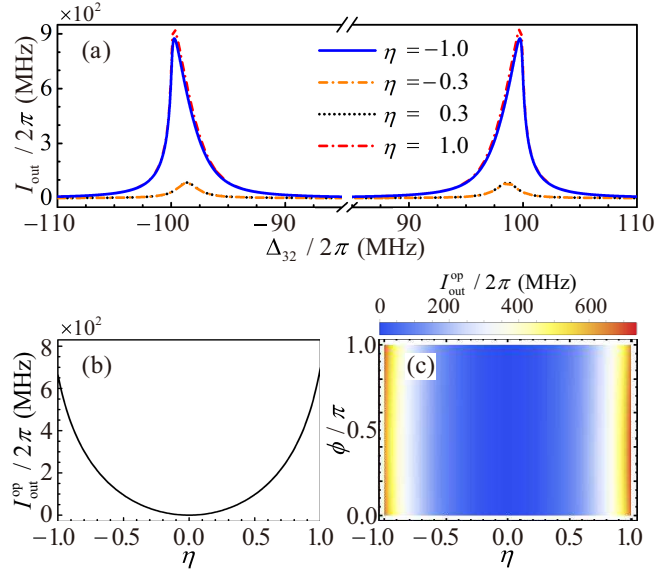}
	\caption{(a) The intensity of the output field as a function of the detuning $\Delta_{32}$ for different enantiomeric excess $\eta$. (b) The optimal cavity output intensity $I_{\rm{out}}^{\rm{op}}$ versus the enantiomeric excess $\eta$. (c) The optimal cavity output intensity $I_{\rm{out}}^{\rm{op}}$ versus the enantiomeric excess $\eta$ and the overall phase $\phi$. Here $\phi=0$ [for panels (a) and (b)], $\Delta_{32} = g_a \sqrt{N}$, $\Omega_{31}/2\pi=20\,\text{kHz}$, and $\Omega_{32}/2\pi=0.5\,\text{MHz}$. The other parameters are the same as those in Fig.~\ref{eeDepenLOW}.}
	\label{eeDepenNLOW}
\end{figure}

Since the enhancement of $\Omega_{31}$ can lead to more molecules occupying the excited states $\left|2\right\rangle$ and $\left|3\right\rangle$, we now take stronger coupling strength $\Omega_{31}$ (e.g. $\Omega_{31}/2\pi \sim 20\,\text{kHz}$) than that considered in the case of low-excitation limit of molecules (e.g., $\Omega_{31}/2\pi=5\,\text{kHz}$). Similar to the case of low-excitation limit of molecules, there are two characteristic peaks for $I_{\rm{out}}$ [see Fig.~\ref{eeDepenNLOW}(a)] located near the detuning $\Delta_{32} \simeq \pm g_{a}\sqrt{N}$. Differently, now the positions of the characteristic peaks vary slightly with $\eta$. For simplicity, we are still interested in the optimal cavity output intensity $I_{\rm{out}}^{\rm{op}}$ at $\Delta_{32} =g_{a}\sqrt{N}$. Note that the optimal cavity output intensity $I_{\rm{out}}^{\rm{op}}$ is enhanced dramatically [e.g., $I_{\rm{out}}^{\rm{op}}/2\pi \simeq 700\,\text{MHz}$ for $\eta=1$ as shown in Fig.~\ref{eeDepenNLOW}(b)]. That means much more photons are generated via the cavity-assisted three-photon processes compared with the case of low-excitation limit of molecules. Therefore, monitoring the optimal cavity output intensity $I_{\rm{out}}^{\rm{op}}$ may provide promising applications in high performance detection of enantiomeric excess in the case beyond low-excitation limit of molecules. Meanwhile, $I_{\rm{out}}^{\rm{op}}$ is different for the enantiopure left- ($\eta=1$) and right- ($\eta=-1$) handed molecules in this case. In addition, we find that the optimal cavity output intensity $I_{\rm{out}}^{\rm{op}}$ also depends on the overall phase $\phi$ [see Fig.~\ref{eeDepenNLOW}(c)]. This is different from the case of low-excitation limit of molecules where $I_{\rm{out}}^{\rm{op}}$ does not depend on $\phi$. For simplicity, we take the overall phase $\phi=0$ in the following discussions.

\begin{figure}[ht]
	\centering
	\includegraphics[width=8.5cm]{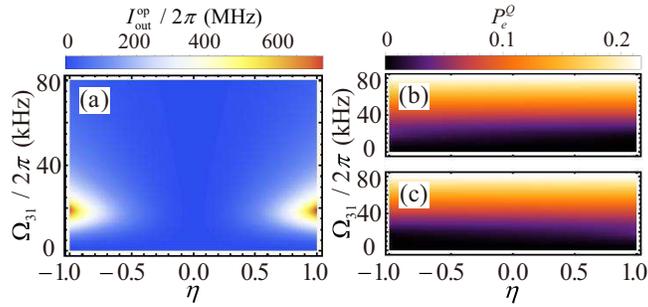}
	\caption{(a) The optimal cavity output intensity $I_{\rm{out}}^{\rm{op}}$, (b) the factor $P_e^L$, and (c) the factor $P_e^R$ versus the enantiomeric excess $\eta$ and the coupling strength $\Omega_{31}$. The other parameters are the same as those in Fig.~\ref{eeDepenNLOW}(b).}
	\label{eeDetectNLOW}
\end{figure}

Further, we show the optimal cavity output intensity $I_{\rm{out}}^{\rm{op}}$ versus the enantiomeric excess $\eta$ and the coupling strength $\Omega_{31}$ in Fig.~\ref{eeDetectNLOW}(a). It is shown that $I_{\rm{out}}^{\rm{op}}$ strongly depends on the coupling strength $\Omega_{31}$. For the coupling strength $\Omega_{31}/2\pi \sim 15\,-\,25\,\text{kHz}$ which is larger than that (e.g., $\Omega_{31}/2\pi = 5\,\text{kHz}$) considered in the case of low-excitation limit of molecules, a certain proportion of the molecules occupies the excited states [see Figs.~\ref{eeDetectNLOW}(b) and~\ref{eeDetectNLOW}(c)], and the cavity-assisted three-photon processes enable the continuous generation of more intracavity photons compared with those in the case of low-excitation limit of molecules. This leads to the dramatic enhancement of the optimal cavity output intensity. Therefore, by adjusting the coupling strength $\Omega_{31}$ appropriately (i.e., $\Omega_{31}/2\pi \simeq 20\,\text{kHz}$), one can obtain the maximal $I_{\rm{out}}^{\rm{op}}$, which improves the resolution in the detection of the enantiomeric excess via measuring the cavity output field.

\section{Conclusion}\label{summary}

In conclusion, we have proposed the method for enantio-detection of chiral molecules in the cavity-molecules system. The key idea is to generate the intracavity photons via the cavity-assisted three-photon processes based on the cyclic three-level models of chiral molecules. In the low-excitation limit of molecules in the presence of weak coupling strength (e.g., $\Omega_{31}/2\pi= 5\,\text{kHz}$), our analytic results show the intensity of the cavity output field in the steady state is proportional to the square of the enantiomeric excess, which provides the method for enantio-detection of chiral molecules by monitoring the cavity output field. Furthermore, by increasing the coupling strength (e.g., $\Omega_{31}/2\pi \sim 20\,\text{kHz}$), we present the results of enantio-detection based on the numerical solution of the cavity output intensity beyond the low-excitation limit of molecules. It is found that the cavity output intensity in the case beyond low-excitation limit of molecules can be enhanced dramatically. This offers the possibility of high performance detection of enantiomeric excess via measuring the cavity output field.

We would like to remark that in the previous methods for enantio-detection~\cite{Barron-MolecularScatter,Discrimination-CD,Discrimination-ChenYY,Discrimination-JSepSci2007,Discrimination-JiaWZ,Discrimination-Lehmann,Discrimination-Nature2000,Discrimination-OR,Discrimination-Science2002,Discrimination-VCD,Discrimination-XuXW,Discrimination-YeC,Discrimination-ZhangXD1,Discrimination-ZhangXD2,Microwave-Doyle-Nature,Microwave-Doyle-PCCP,Microwave-Doyle-PRL,Microwave-Lehmann-JPCL,Microwave-Schnell-ACIE,Microwave-Schnell-JPCL}, the single-molecule treatment was usually used since only the classical light fields are involved. However, in our cavity-molecules system, the chiral molecules confined in the cavity are collectively coupled to the quantized cavity field. Therefore, we should resort to the multi-molecule treatment~\cite{CQEDMolecule2,CQEDMolecule4,CQEDReaction1,CQEDtransfer2} since the single-molecule one is no longer valid in the present cavity-molecules system.

Note that the phase matching is still an important issue~\cite{PhaseMatching-Lehmann-2018,Discrimination-Lehmann} in the methods for enantio-detection based on the cyclic three-level models of chiral molecules~\cite{Microwave-Doyle-Nature,Microwave-Doyle-PCCP,Microwave-Doyle-PRL,Microwave-Lehmann-JPCL,Microwave-Schnell-ACIE,Microwave-Schnell-JPCL}. In these models, the propagation directions of the three electromagnetic fields cannot be parallel~\cite{Seperation-Koch-JCP}. Usually, this would lead to the phase-mismatching problem~\cite{PhaseMatching-Lehmann-2018,Discrimination-Lehmann} due to the finite size of the sample. Moreover, the coupling strength $g_a$ between the molecules and the quantized cavity field will be space-dependent due to the spatial distribution of the sample along the cavity axis. To ensure that all the molecules are approximately phase matched and the coupling strength $g_a$ is approximately identical for each molecule, the sample is required to be fixed in a volume whose size is much smaller than the typical wavelengths of the applied electromagnetic fields. With the further development of molecular cooling technology~\cite{MolecularCooling-McCarron-2018,MolecularCooling-Augenbrau-2020}, it is expected to prepare cold sample with smaller size in experiments so that the influence of phase-mismatching and spatial dependence of coupling strength $g_a$ can be negligible.

\section*{Funding}
Science Challenge Project (TZ2018003); National Natural Science Foundation of China (11774024,
12074030, U1930402); National Science Foundation for Young Scientists of China (12105011); Beijing Institute of Technology Research Fund Program for Young Scholars.

\section*{Disclosures}
The authors declare that there are no conflicts of interest related to this article.

\section*{Data availability}
Data underlying the results presented in this paper are not publicly available at this time but may be obtained from the authors upon reasonable request.


\end{document}